\documentclass[iop,preprint]{emulateapj}
\usepackage{graphicx}
\usepackage{amsmath}
\usepackage{natbib}

\begin{document}

\slugcomment{submitted to AJ}
\shortauthors{Harris}

\title{Where Are Most of the Globular Clusters in Today's Universe?}
        
\author{William E. Harris\altaffilmark{1}
}

\altaffiltext{1}{Department of Physics \& Astronomy, McMaster University, Hamilton, ON, Canada; harris@physics.mcmaster.ca}

\date{\today}

\begin{abstract}
	The total number of globular clusters (GCs) in a galaxy rises continuously with the 
	galaxy luminosity $L$, while the relative number of galaxies decreases with $L$ 
	following the Schechter function.  The product of these two very nonlinear functions
	gives the relative number of GCs contained by all galaxies at a given $L$.
	It is shown that GCs, in this universal sense, are most commonly found in galaxies
	within a narrow range around $L_{\star}$.  In addition, blue (metal-poor) GCs outnumber
	the red (metal-richer) ones globally by 4 to 1 when all galaxies are added, pointing to the 
	conclusion that the earliest stages of galaxy formation were especially favorable to
	forming massive, dense star clusters.
\end{abstract}


\section{Introduction}

Galaxies with luminosities higher than $L \sim 10^7 L_{\odot}$ -- in essence, all but the very smallest dwarfs -- 
have measurable numbers
of globular clusters (GCs), the massive compact star clusters that were preferentially formed during
the earliest stages of star formation during galaxy evolution.  The number of GCs
present in a given galaxy increases dramatically with host galaxy mass or luminosity, but not in a simple linear way
\citep[][hereafter HHA13]{harris_etal13}.  At the same time, the relative number of galaxies decreases
continuously with $L$ following the empirically based Schechter function \citep{schechter1976}.    

Combining these two opposing trends leads to a rather simple question:  which galaxies contribute the most
to the total number of globular clusters in the universe?  Dwarf galaxies have very few GCs 
individually, but there are huge numbers of such galaxies.  Contrarily, the biggest GC populations 
are to be found in central supergiant ellipticals like M87, but these are very rare galaxies.  Which ones are the
most important when added up over the entire galaxy population?  

At the same time, we can address the question of the two classic subpopulations of GCs, the
blue (metal-poor) and red (metal-rich) ones that are consistently seen to form a bimodal distribution
in GC luminosity versus color \citep[e.g.][]{brodie_strader06}.  
Color index increases monotonically with GC metallicity and thus is
a useful proxy for [Fe/H], with the dividing line between blue and red near [Fe/H] $\simeq -1$.
Whereas the blue GCs are consistently found in
all galaxies from dwarfs to giants, the red ones reside preferentially in massive galaxies;
quantitative discussions of this trend are given by, e.g., \citet{peng_etal06,peng_etal08} and \citet{harris_etal2015} 
(hereafter HHH15).
Because the metal-poor blue GCs are found in all galaxies, it could therefore be expected that they 
would outnumber the metal-richer ones in total, but it is not immediately clear by how much.  

In this paper, some simple GC demographics are calculated to gain
a first answer to these questions.  As will be seen below, the discussion draws heavily on recent observational gains
that establish the numbers of blue, red, and all GCs within galaxies covering their entire luminosity
range (see HHA13, HHA15).  In what follows I have adopted a distance scale of $H_0 = 70$ km s$^{-1}$ Mpc$^{-1}$
wherever necessary.

\section{Analysis}

\begin{figure}[t]
	 \vspace{0.0cm}
	  \begin{center}
		   \includegraphics[width=0.5\textwidth]{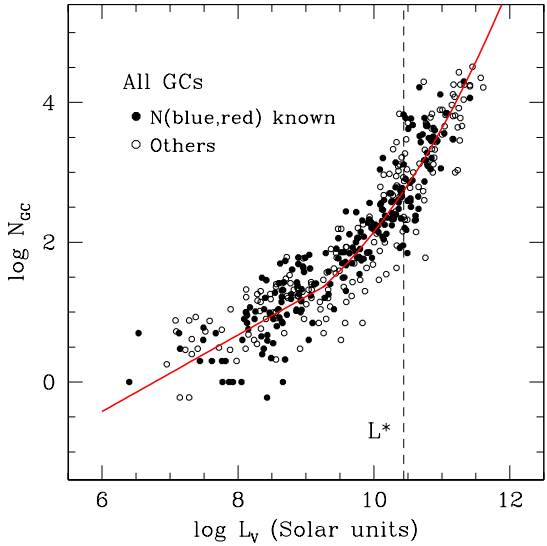}
	   \end{center}
	   \vspace{-0.5cm}
	   \caption{Total number of globular clusters $N_{GC}$ versus host galaxy luminosity $(L_V/L_{\odot})$,
		   with data from the catalog of HHA13.  \emph{Solid symbols} denote galaxies in which the relative numbers of
		   blue (metal-poor) and red (metal-rich) GCs are known, while \emph{open symbols} denote galaxies in which only
           the totals have been estimated. The equation for the interpolation curve is given in the text.  The vertical 
   dashed line shows $L_{\star}$ from the Schechter function.}
	   \vspace{0.5cm}
	   \label{fig:ngctot}
   \end{figure}

In Figure \ref{fig:ngctot}, the total number of GCs ($N_{GC}$) is plotted versus host galaxy luminosity
($L$), from data for 418 galaxies of all
types as listed in the recent catalog of HHA13.  Here, no discrimination is made by galaxy type (spiral, S0, E),
but as shown in HHA13 and HHH15, differences by type appear to have only second-order effects.  
For about half the observed sample (n=216), 
the original observations are of sufficient photometric precision
and depth to resolve the standard bimodal distribution in GC colors and thus to obtain the red and blue fractions as well.

For our purposes here, a useful interpolation curve giving the trend of $N_{GC}$ versus $L(gal)$, is
\begin{equation}
	{\rm log} N_{GC} = 
	\begin{cases}
		-3.71 + 0.548 x  & (x < 9.35), \\
		-0.30 + 0.66 x - 0.1815 x^2 + 0.014 x^3 & (x \geq 9.35)
	\end{cases}
\end{equation}
\noindent where $x$ = log $(L_V/L_{\odot})$.  A marked change in the slope of this relation happens
near $L \sim 2 \times 10^9 L_{\odot}$; for the dwarfs fainter than that transition we find
$N_{GC} \sim L^{0.5}$, while for large galaxies above it we find roughly $N_{GC} \sim L^{1.4}$.
The ratio of these two quantities is the classic specific frequency $S_N = const (N_{GC}/L_V)$
\citep{harris_vandenbergh1981}, which has a well known characteristic U-shaped dependence on $L$.

\begin{figure}[t]
	 \vspace{0.0cm}
	  \begin{center}
		   \includegraphics[width=0.5\textwidth]{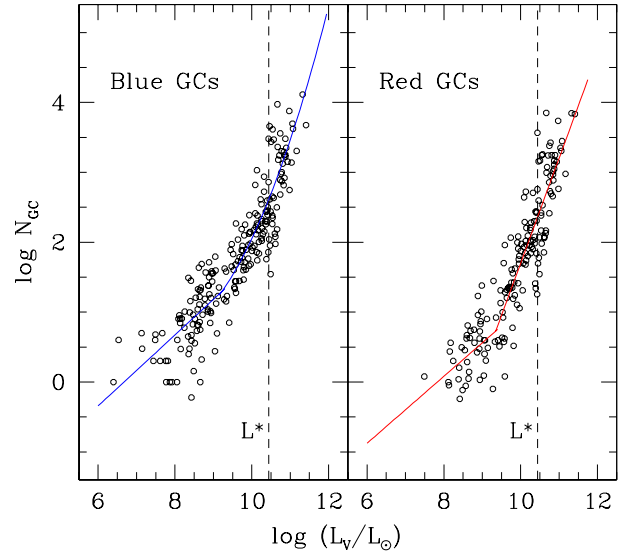}
	   \end{center}
	   \vspace{-0.5cm}
	   \caption{\emph{Left panel:}  Total number of blue (metal-poor) globular clusters 
		   versus host galaxy luminosity; \emph{Right panel:} total number of red
		   (metal-rich) clusters.  The equations for the interpolation curves are given in the text.  
	   The vertical dashed lines denote the Schechter-function $L_{\star}$.}
	   \vspace{0.5cm}
	   \label{fig:redblue}
   \end{figure}

In Figure \ref{fig:redblue}, the same relation is shown but now divided into the blue and red
subpopulations.  The approximate interpolation curves shown in the Figure are, for the blue GCs,
\begin{equation}
	{\rm log} N_{GC} = 
	\begin{cases}
		-3.39 + 0.51 x  & (x < 9.35), \\
		15.418 - 3.84 x + 0.25 x^2  & (x \geq 9.35)
	\end{cases}
\end{equation}
\noindent and for the red GCs,
\begin{equation}
	{\rm log} N_{GC} = 
	\begin{cases}
		-3.75 + 0.48 x  & (x < 9.35), \\
		-13.31 + 1.50 x  & (x \geq 9.35) \, .
	\end{cases}
\end{equation}

The Schechter function giving the relative number of galaxies per unit luminosity is
\begin{equation}
	\phi(L) dL \, = \, \phi_o ({L \over L_{\star}})^{\alpha} e^{-(L/L_{\star})} dL \, .
\end{equation}
\noindent The parameters ($\alpha, L_{\star}$) may empirically depend somewhat on environment, but
in this case the goal is simply to track the first-order behavior of GC populations
averaged over all environments.
The values adopted here are $\alpha = -1.26$ and $L_{\star} = 2.77 \times 10^{10} L_{\odot}$
from the SDSS DR6 database discussed by \citet{montero-dorta2009}; many other versions close
	to this pair of values can be found in the recent literature, but the precise numbers do not
	affect the results of the following discussion in any significant way.

\begin{figure}[t]
	 \vspace{0.0cm}
	  \begin{center}
		   \includegraphics[width=0.5\textwidth]{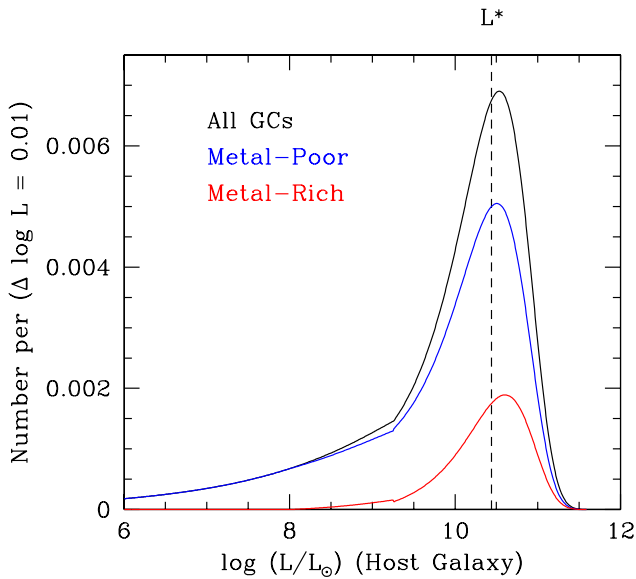}
	   \end{center}
	   \vspace{-0.5cm}
	   \caption{Relative number of globular clusters in all galaxies at a given luminosity (log $L$).  
		   The numbers are for a constant bin size $\Delta {\rm log} L = 0.01$ ($\Delta M_V = 0.025$ magnitude).
	   The values are normalized so that the total integrated over all $L$ equals 1.  \emph{Black curve:} 
   the distribution for all GCs combined, regardless of metallicity; \emph{Blue curve:} the distribution for only the 
   metal-poor (blue) GCs; and \emph{Red curve:} the distribution for only the metal-rich (red) GCs.  The sum
   of the blue and red curves by definition equals the upper curve.  The vertical dashed line denotes the
   Schechter-function $L_{\star}$.}
	   \vspace{0.5cm}
	   \label{fig:nbin}
   \end{figure}

Calculating the total number of GCs in all galaxies at a given $L$ is then a matter of multiplying
Equation (4) numerically with either (1), (2), or (3) depending on which GC subpopulation we want to track.  
The results in smoothed histogram form are shown in Figure \ref{fig:nbin}, which gives the total number of
GCs in all galaxies within a constant \emph{logarithmic} bin size $\Delta {\rm log} L = 0.01$.
In rough terms, this graph gives the relative probability that a globular cluster anywhere in the universe
will be sitting in a host galaxy of luminosity (log $L$), or equivalently at a given absolute magnitude.

The shapes of all three curves in Fig.~\ref{fig:nbin} peak strongly at intermediate luminosities very near
$L_{\star}$, with a long gradual ramp down towards the dwarf galaxies at lower $L$.  Dwarf
galaxies are very common but they do not have enough GCs per galaxy to dominate the totals; and contrarily,
the highest$-L$ supergiant ellipticals have tens of thousands of GCs each but they are too rare to dominate.  
What is perhaps surprising is the height and relative sharpness of the population peak.  
Specifically, we find the following features:
\begin{itemize}
	\item{} For all GCs combined, the peak is at log $(L/L_{\odot}) \simeq 10.53$ and 50\% of the population
		lies in the range log $(L/L_{\odot}) = 9.86 - 10.95$ (a factor of 12 in $L$).
	\item{} For the blue GCs, the peak is at log $(L/L_{\odot}) \simeq 10.5$ and 50\% of the population
		lies between log $(L/L_{\odot}) = 9.78 - 10.93$ (a factor of 14).
	\item{} For the red GCs, the peak is at log $(L/L_{\odot}) \simeq 10.6$ and 50\% of them lie
		between log $(L/L_{\odot}) = 10.03 - 10.99$ (a factor of 9).
\end{itemize}

The reason why these peaks are rather high and narrow can be seen from Figs.~\ref{fig:ngctot} and \ref{fig:redblue}.
$N_{GC}$ begins rising steeply near log $(L/L_{\odot}) \sim 9.5$, which is still a decade
below $L_{\star}$.  Thus the 
Schechter function is still on the flat part of its curve ($L < L_{\star}$) and the number of galaxies
at a given $L$ is declining only slowly.  Once $L$ passes $L_{\star}$, however, the number of host
galaxies declines so steeply that it forces all the curves in Fig.~\ref{fig:nbin} rapidly downward.
In short, the galaxies near $L_{\star}$ provide the ``best compromise'' situation for GC populations
in a universal sense:  they have typically several hundred GCs per galaxy, and are still numerous enough
cosmologically to dominate the GC totals.

The general appearance of Fig.~\ref{fig:nbin} to some extent resembles the \citet{li_white2009}
model calculation of the total amount of stellar mass contributed by galaxies of a given luminosity
or baryonic mass (see their Fig.~5).  In both cases the distribution is sharply peaked near 
$\L_{\star}$.  However, the long tail towards low $L$ for the GC numbers is noticeably more
prominent than for all stellar mass, reflecting the empirical fact that dwarf galaxies have
higher average specific frequencies than $L_{\star}$-type ones.

\begin{figure}[t]
	 \vspace{0.0cm}
	  \begin{center}
		   \includegraphics[width=0.5\textwidth]{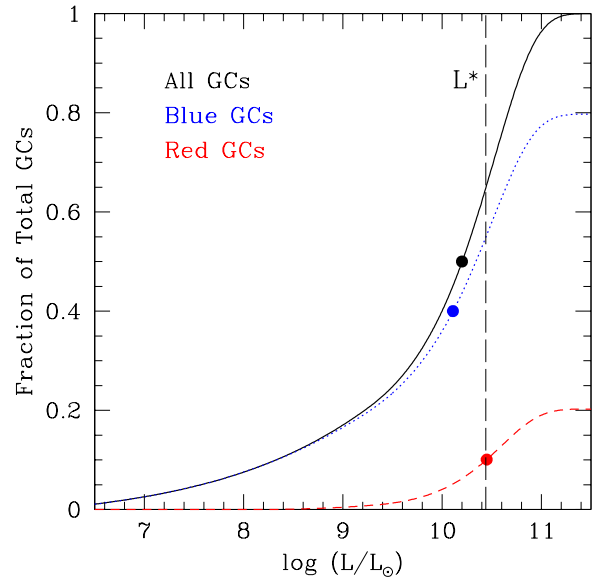}
	   \end{center}
	   \vspace{-0.5cm}
   \caption{The fraction of all globular clusters within galaxies with luminosities $\leq L$.
   The large solid dots indicate the halfway points along each curve; i.e. half the clusters are
   found in galaxies with $L \leq L_{\bullet}$ in each case.  The vertical dashed line denotes the
   Schechter-function $L_{\star}$.}
	   \vspace{0.5cm}
	   \label{fig:ntot}
   \end{figure}

For comparison, Figure \ref{fig:ntot} presents the cumulative distribution.  Half the population of blue GCs
resides in galaxies with $L < 1.3 \times 10^{10} L_{\odot}$, whereas half the red GC population falls within
galaxies with $L < 2.8 \times 10^{10} L_{\odot}$, a crossing point more than twice as high.

It is also noteworthy from Fig.~\ref{fig:ntot} 
that the blue, metal-poor GCs make up almost 80\% of all globular clusters in the universe.
A major reason for this predominance is that in the dwarf-galaxy regime (log $(L/L_{\odot} \lesssim 9.5$)
there are almost no metal-rich GCs present, and it is only in the very luminous (and rare) supergiants
that they make up comparable numbers to the metal-poor ones
\citep[see, e.g.,][for recent examples]{peng_etal06,harris09a,harris_etal2014}.  

\begin{figure}[t]
	 \vspace{0.0cm}
	  \begin{center}
		   \includegraphics[width=0.5\textwidth]{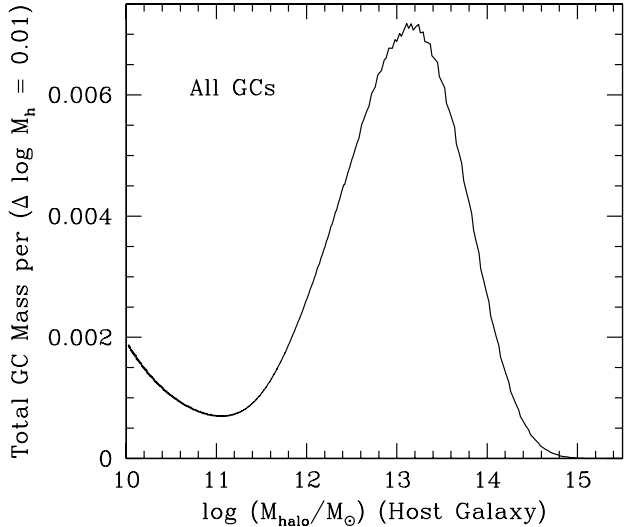}
	   \end{center}
	   \vspace{-0.5cm}
	   \caption{Total mass in all GCs within galaxies at a given halo mass $M_h$.
		   The numbers are for a constant bin size $\Delta {\rm log} M_h = 0.01$.
	   The values are normalized so that the total integrated over all $L$ equals 1.  \emph{Black curve:} 
   the distribution for all GCs combined, regardless of metallicity; \emph{Blue curve:} the distribution for only the 
   metal-poor (blue) GCs; and \emph{Red curve:} the distribution for only the metal-rich (red) GCs.  The sum
   of the blue and red curves by definition equals the upper curve.  The vertical dashed line denotes the
   Schechter-function $L_{\star}$.}
	   \vspace{0.5cm}
	   \label{fig:nbin_mh}
   \end{figure}

In Figure \ref{fig:nbin_mh}, another version of the probability distribution is shown, but now
plotted versus galaxy halo mass $M_h$ rather than luminosity; $M_h$ is dominated by dark matter.  
The Figure shows the total \emph{mass} in GCs within
all galaxies at a given $M_h$ rather than the total \emph{number}, but these
are nearly equivalent given the very shallow increase of mean GC mass with galaxy mass (HHA13).
This graph was generated through the combination of 
(a) a double-Schechter-function form of the number of galaxies per unit stellar mass $M_{\star}$,
from \citep{kelvin_etal2014},
(b) conversion of $M_{\star}$ to $M_h$ via the stellar-to-halo mass ratio SHMR with the
convenient parametrization of \citet{guo_etal2010}, and finally
(c) the total mass in GCs within a galaxy of a given $M_h$, which has a simple linear form (HHH15).

The graph indicates that GCs are most likely to be found within galaxy halos near $\sim 10^{13} M_{\odot}$,
but the peak is much broader than in Fig.~\ref{fig:nbin}, 
a result of the very nonlinear conversion of $M_{\star}$ (baryonic mass) 
to $M_h$.  The slight upturn
of the curve for $\lesssim 10^{11} M_{\odot}$ is quite uncertain (see HHH15 for a discussion of the data), but is
partly determined by the steeper slope of the double Schechter function for the smallest dwarfs
\citep{kelvin_etal2014}.  However, this
calculation is more or less arbitrarily cut off below $M_h = 10^{10} M_{\odot}$, since 
dwarfs below this limit have $< 1$ GC each according to the empirical evidence (see HHH15).

\section{Summary and Discussion}

In this paper some broad-brush demographics of globular cluster populations are discussed; for
the first time, it is possible to estimate quantitatively (though admittedly only to first order) which
galaxies are responsible for contributing most of the GCs in the present-day universe.  The combination
of the nonlinear shapes of both the $N_{GC}$ versus $L$ function, and the Schechter function
for galaxy numbers, demonstrates that galaxies in a narrow range around the Schechter $L_{\star}$ contribute the most.  

Expressed in terms of galaxy halo mass (i.e. total mass) rather than luminosity, GCs are predominantly found 
within halos in the broad range $\sim 10^{12-14} M_{\odot}$, with the peak near $10^{13} M_{\odot}$.

The primary result of this discussion is shown in Fig.~\ref{fig:nbin}.  It should be seen essentially as a snapshot in 
time, valid only for the present day:  as the universe evolves and the continual process of galaxy merging
continues, the biggest galaxies grow by absorbing their small neighbors.  Thus over time, the
peak in Fig.~\ref{fig:nbin} will shift to higher $L$ and the tail at lower $L$ will shrink.  
In the past at much higher redshift, the galaxy population was much more dominated by dwarfs and the 
GC population peak was correspondingly at lower $L$.

The result that the metal-poor GCs outnumber the metal-richer ones by a global ratio of roughly 4 to 1
is striking.  The implication for galaxy evolution is that the very earliest stages of hierarchical merging,
when baryonic matter was predominantly in quite low-metallicity gas, was exceptionally favorable for
the formation of dense massive star clusters \citep[see also][]{kruijssen2015}.  The bulk of low-metallicity
GC formation appears to have happened near $z \sim 5$ \citep[e.g.][and references cited there]{vandenberg_etal13,forbes_etal2015},
while the metal-richer ``red'' population predominated later near $z \sim 2-3$, much nearer the peak of
the cosmic star formation rate \citep[e.g.][]{madau_dickinson2014}.  At that later time, the remaining gas 
was much more enriched, but it was not
as successful at producing the extremely dense $\sim 10^5-10^7 M_{\odot}$ protocluster clouds within which GCs could form
\citep{harris_pudritz1994,kravtsov_gnedin05,elmegreen2012,li_gnedin14,kruijssen2015}.

In the discussion above, the term ``globular cluster'' is taken implicitly to mean classically old, massive star clusters.
If we broaden that definition to include massive star clusters formed at any time, then it would be
appropriate to include YMCs (young massive star clusters, also sometimes referred to as super star clusters
in the literature) formed in 
low-redshift mergers between galaxies, as is seen in nearby active merger remnants
\citep[e.g.][among others]{trancho_etal2014,trancho_etal2007,whitmore_etal2014,goudfrooij2012,zepf_etal1999,carlson_etal1998}.
These young GCs will add to the metal-rich GC population and to some extent increase their fraction
of the universal population.  However, even in the most prominent mergers (see the citations above
for examples) only some dozens of clusters are added that are $\gtrsim 10^5 M_{\odot}$ and thus likely
to survive for many Gyr.  These will not add significantly to the old clusters already present when
added up over all galaxies.  In essence, the GC production in any merger old or young will depend
critically on the amount of cold gas available, so most GC formation happened in the early universe
\citep[see][for discussion]{li_gnedin14}.

Lastly, the present discussion does not account for GCs either in the Intergalactic Medium (IGM)
or Intracluster Medium (ICM), i.e. ones not definitely associated with any individual galaxy.  True
IGM clusters far from any galaxy are extremely hard to find and their numbers are generally
presumed to be be very small, lacking any evidence to the contrary.  
ICM populations of GCs are also not well studied as yet, but are known
to exist in a few rich clusters of galaxies such as Virgo or Coma 
\citep{peng_etal11,durrell_etal2014,west_etal2011,alamo-martinez_etal2013}.  The cases studied so far indicate
that these ICM GCs add up to roughly the same numbers as are associated with the central Brightest Cluster
Galaxy in their local environment, and would therefore not change the totals estimated here significantly.

\section*{Acknowledgements}

The author acknowledges financial support from NSERC (Natural Sciences and Engineering Research Council of Canada).

\makeatletter\@chicagotrue\makeatother

\bibliographystyle{apj}

\%label{lastpage}

\end{document}